\documentclass[utf8]{FrontiersinHarvard} 

\usepackage{url,hyperref,microtype,subcaption}
\usepackage[onehalfspacing]{setspace}

\def\rfe{R$_{\rm Fe II}$}

\def\feii{Fe{\sc ii}}

\def\keyFont{\fontsize{8}{11}\helveticabold }
\def\firstAuthorLast{Panda} 

\def\Address{$^{1}$ International Gemini Observatory/NSF NOIRLab, Casilla 603, La Serena, Chile\\
$^{2}$ Laborat\'orio Nacional de Astrof\'isica - MCTI, R. dos Estados Unidos, 154 - Na\c{c}\~oes, Itajub\'a - MG, 37504-364, Brazil\\

$^{\ddagger}$ Gemini Science Fellow}
\def\corrAuthor{Corresponding Author}

\def\corrEmail{swayamtrupta.panda@noirlab.edu}

\begin{document}
\onecolumn
\firstpage{1}

\title[Quasar Main Sequence: current state and recent advances]{Unveiling The Quasar Main Sequence: Illuminating The Complexity Of Active Galactic Nuclei And Their Evolution} 

\author{Swayamtrupta Panda\,$^{1,2,^{\ddagger,}*}$} 
\address{} 
\correspondance{} 

\extraAuth{}

\maketitle

\begin{abstract}

The Eigenvector 1 schema, or the main sequence of quasars, was introduced as an analogous scheme to the HR diagram that would allow us to understand the more complex, extended sources - active galactic nuclei (AGNs) that harbor accreting supermassive black holes. The study has spanned more than three decades and has advanced our knowledge of the diversity of Type-1 AGNs from both observational and theoretical aspects. The quasar main sequence, in its simplest form, is the plane between the FWHM of the broad H$\beta$ emission line and the strength of the optical \feii{} emission to the H$\beta$. While the former allows the estimation of the black hole mass, the latter enables direct measurement of the metal content and traces the accretion rate of the AGN. Together, they allow us to track the evolution of AGN in terms of the activity of the central nuclei, its effect on the line-emitting regions surrounding the AGN, and their diversity making them suitable distance indicators to study the expansion of our Universe. This mini-review aims to provide (i) a brief history leading up to the present day in the study of the quasar main sequence, (ii) introduce us to the many possibilities to study AGNs with the main sequence as a guiding tool, and (iii) highlight some recent, exciting lines of researches at the frontier of this ever-growing field.

\tiny
 \keyFont{ \section{Keywords:} active galactic nuclei; quasars; Seyfert galaxies; emission lines; AGN variability; changing-look AGNs; accretion disks} 
\end{abstract}

\section{What is the Quasar Main Sequence? - a brief history}

More than three decades ago, \citet{boroson_green_1992} put forward the idea of a main sequence of quasars - an analogous schema to the Hertzsprung-Russel (HR) diagram \citep{hertszsprung_1911, Russell_1914} that has allowed us to track the evolution of stars of varied ages and diverse properties utilizing the classification based on their color and magnitude. Akin to the HR diagram, the quasar main sequence (QMS) was envisioned to help put together the diverse population of Type-1, unobscured active galactic nuclei (AGNs) through the compilation of spectral properties from the broad- and narrow line-emitting regions of a sample of nearby, bright AGNs.

Before diving into the recent advances, we would like to reflect on the importance of \citeauthor{boroson_green_1992}'s work with a brief account of the procedure carried out to realize the first results that laid the foundations of the Quasar Main Sequence.

\subsection{The inception of the Main Sequence of quasars}

\begin{figure}[htb!]
    \centering
    \includegraphics[width=0.495\linewidth]{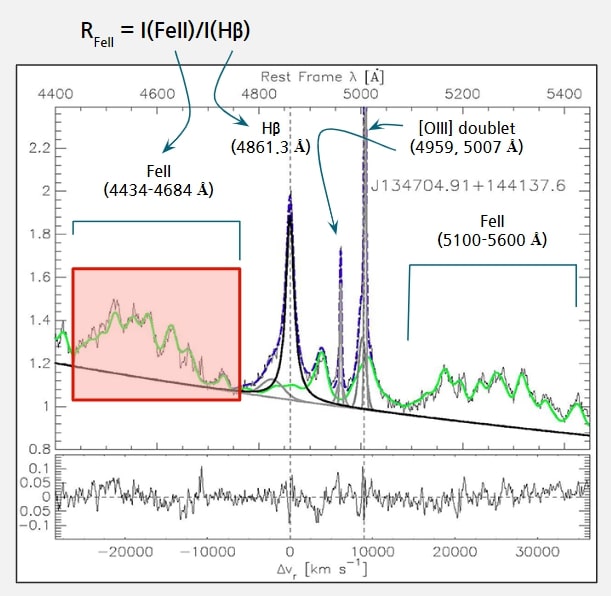}
    \includegraphics[width=0.46\linewidth]{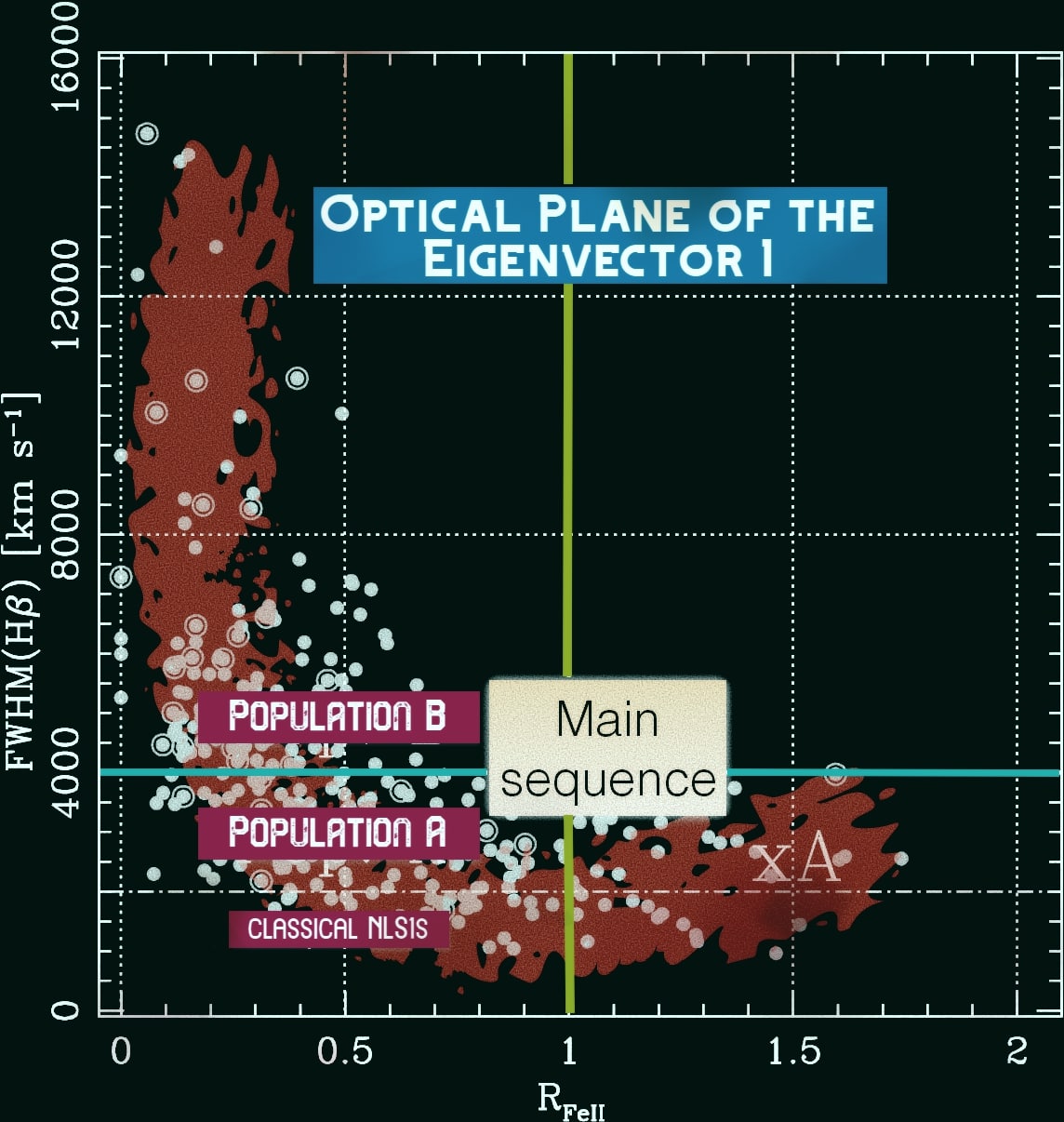}
    \caption{\textbf{Left:} Spectral decomposition (optical region) of a Type-1 Narrow-line Seyfert (NLS1) galaxy, SDSS J134704.91+144137.6. The original spectrum is shown in light gray, the H$\beta$ profile is fit with a Lorentzian function (solid black) and a blue-shifted outflowing component (solid grey), the fit to the H$\beta$-[O {\sc iii}] complex is shown in dashed gray, and in green the \feii{} pseudocontinuum fit is shown. The shaded region highlights the \feii{} blend within 4434-4684 \AA\ used to estimate the \feii{} strength (wrt broad H$\beta$), i.e., \rfe{}. The residua from the fit is shown in the bottom panel. Credit: \citet{negrete_etal_2017FrASS}; \textbf{Right:} A Schematic diagram of the optical plane of Eigenvector 1. The solid horizontal line (turquoise) represents the threshold in FWHM(H$\beta$) at 4000 km s$^{-1}$, which distinguishes between Population A and Population B sources \citep{marziani_etal_2018}. The ``classical'' NLS1s are situated below the FWHM(H$\beta$) $\lesssim$ 2000 km s$^{-1}$ mark (indicated by the dotted-dashed line). The vertical green line marks the boundary for \rfe{} = 1, which separates weak from strong \feii{} emitters, also known as xA sources. Credit: \citet{panda_etal_2024}, \citet{marziani_etal_2018}}
    \label{fig:enter-label}
\end{figure}

\citeauthor{boroson_green_1992} conducted their study within the low-redshift range (z $<$ 0.5), analyzing 87 sources from the Bright Quasar Survey \citep{schmidt_green_1983}. Their primary finding from optical spectra analysis of these quasars was that the \feii{} line equivalent width consistently matched that of H$\beta$, indicating that \feii{} emission originates from the same broad-line region (BLR) clouds as H$\beta$. Additionally, they compiled optical spectral properties for each source. They integrated these with data from other spectral regions from previous studies, creating a 17-parameter correlation matrix for emission lines and continuum properties. Using principal component analysis (PCA, \citealt{Francis_1999ASPC}) on this matrix to identify meaningful correlations, they focused on 13 key properties including $M_{V}$ (V-band magnitude), log $R$ (radio-to-optical spectral index, \citealt{Kellermann_1989AJ}), $\alpha_{ox}$ (optical-to-X-ray spectral index, \citealt{Tananbaum_1986ApJ}), EW(H$\beta$), [O{\sc iii}]$\lambda$5007 strength, He{\sc ii}$\lambda$4686 strength, \feii{} (4434-4684\AA) strength (see also left panel of Figure \ref{fig:enter-label}), [O{\sc iii}]$\lambda$5007/H$\beta$ peak height ratio, FWHM(H$\beta$), H$\beta$ profile shift, shape, asymmetry, and $M_{[O{\sc iii}]}$. They determined that the primary eigenvector (Eigenvector 1 or EV1) derived from these properties was primarily characterized by the anti-correlation between \feii{} strength (specifically \rfe{}, the ratio of EW(\feii{}) for the 4434-4684\AA\ blend to EW(H$\beta$) and [O{\sc iii}]$\lambda$5007 strength. EV1 also showed significant correlations ($>|0.5|$) with log $R$, FWHM(H$\beta$), and H$\beta$ profile asymmetry.

To interpret the PCA results, the authors identified several key parameters that could influence the observed properties: (1) the mass accretion rate, (2) the black hole (BH) mass, (3) the covering factor of the BLR clouds, (4) the degree of anisotropy in the emitted radiation from the continuum source, (5) the orientation of the source to the observer, (6) the velocity distribution of the BLR clouds, and (7) the ionization parameter.

To summarize, the paper by \citet{boroson_green_1992} is fundamental for two main reasons:

1. It is one of the first publications in AGN research to use principal component analysis (PCA) to explore the connections between the observed properties of quasars, particularly in the study of the Quasar Main Sequence (see right panel of Figure \ref{fig:enter-label} for a recent rendition). This sequence unifies the diverse group of AGNs through Eigenvectors, specifically, Eigenvector 1, which shows an anti-correlation between the width of the optical \feii{} blend (4434-4684\AA) and the peak intensity of the forbidden [OIII]$\lambda$5007\AA\ line. The study also established a connection between the width of the broad H$\beta$ emission and this eigenvector, forming the well-known ``Quasar Main Sequence", primarily driven by the Eddington ratio among other physical properties \citep[e.g.,][and references therein]{sulentic_etal_2000, shen_ho_2014, marziani_etal_2018, panda_etal_2019ApJ_QMS}.

2. For the first time, the paper constructed the \feii{} pseudo-continuum template from the optical spectrum of I Zw 1. This template has become widely used in analyzing the optical spectra of AGNs, facilitating the study of the \feii{} complex \citep{Phillips_1978ApJS} both theoretically and observationally. It helped understand the excitation mechanisms \citep{Phillips_1978ApJ, Verner_1999ApJS} behind the thousands of spectral transitions from the UV to the NIR, transforming \feii{} from being considered a spectral contaminant to an evolution tracer and fundamental component of the BLR in AGNs \citep{Marinello_2016ApJ, marziani_etal_2018, panda_etal_2019ApJ_QMS, mlma_etal_2021ApJ, panda_2022FrASS}.

\subsection{The broad contextualization in the form of 4D Eigenvector 1}

An advanced version of the Eigenvector 1 (EV1) schema was introduced by \citet{sulentic_etal_2000}, incorporating additional parameters beyond the initial (1) FWHM of the broad component of H$\beta$ and (2) the equivalent width ratio of the optical \feii{} blend (4434-4684 \AA) to broad H$\beta$ (\rfe{}). These new parameters include (3) the centroid shift at FWHM of the high ionization line C IV $\lambda$1549, c(1/2), and (4) the soft X-ray photon index ($\Gamma_{\rm soft}$). Simplified, these measures: (1) the extent of virialized motions in a low-ionization line-emitting accretion disk or a flattened cloud distribution, acting as a virial estimator of black hole mass \citep[e.g.,][]{collin-souffrin_1988MNRAS, dultzin-hacyan_1999ASPC, joly_etal_2008RMxAC} (2) the ionization and size of the BLR cloud, with the \feii{} emission strength (\rfe{}) suggesting its origin near the accretion disk; (3) indicators of winds/outflows in high ionization broad line gas; and (4) thermal emission related to the accretion disk and SMBH accretion state \citep[e.g.,][]{mineshige_etal_2000PASJ, done_etal_2012MNRAS}. We refer the readers to a comprehensive summary in \citet{marziani_etal_2018}.

Accumulating evidence from subsequent studies, following \citet{boroson_green_1992}, indicates that EV1 correlations involve at least two principal independent parameters: (1) the source's bolometric luminosity (L$_{\rm bol}$) and its black hole mass (M$_{\rm BH}$), convolved with source orientation \citep{marziani_etal_2001ApJ, panda_etal_2019ApJ_QMS}. These two parameters are succinctly expressed as the Eddington ratio (L$_{\rm bol}$/L$_{\rm Edd}$).

\subsection{Getting the ``bigger" picture}

The onset of the new century saw the rise of large spectroscopic surveys, such as the Sloan Digital Sky Survey \citep[SDSS,][]{york_etal_2000AJ, shen_etal_2011}. These surveys revitalized EV1 studies and extended their applicability to much larger samples. \citet{shen_ho_2014} made significant strides with their seminal paper, utilizing data from over 20,000 spectroscopically observed SDSS quasars, analyzed using an automated spectral fitting pipeline \citep{shen_etal_2011}. This study provided spectral parameters for a wide range of emission lines, as well as estimates for black hole masses and Eddington ratios. Leveraging this comprehensive dataset, \citeauthor{shen_ho_2014} redefined the main sequence of quasars and concluded that (1) the average Eddington ratio increases from left to right on the sequence, and (2) the dispersion in FWHM(H$\beta$) at a fixed \rfe{} is largely due to orientation effects (see also \citealt{sun_shen_2015}). They proposed that quasar properties correlated with EV1 can be unified by variations in the average Eddington ratio of the accreting black hole, driven by systematic changes in the shape of the accretion disk continuum and its role in photoionizing the line-emitting regions.

More recently, the exploration of large samples has been extended to include sources from the Southern Hemisphere \citep[][and references therein]{chen_etal_2018A&A}, and the number of Type-1 AGNs, including strong \feii{}-emitting ones, has grown many folds with deeper surveys extending to fainter magnitudes \citep{rakshit_etal_2020ApJS, wu_shen_2022ApJS, paliya_etal_2024MNRAS, panda_etal_2024}. We are now at a stage where AGNs are frequently revisited and thus we also have a wealth of multi-epoch, multi-wavelength data for samples of AGNs. This has greatly helped to build samples of AGNs that demonstrate changes in their continuum and emission line properties - the Changing-Look AGNs \citep[see recent compilations in][and references therein]{panda_sniegowska_2024, Guo_DESI_2024, Zeltyn_etal_2024ApJ}, especially investigating the changes in the \feii{} emission in the context of the quasar main sequence \citep{panda_sniegowska_2024}.

The paper is organized as follows: In Section \ref{sec2}, we highlight some recent advances in the last decade on the studies with the quasar main sequence - \feii{} template creation and improvements, theoretical predictions, and advancements in photoionization modeling including some direct confirmations of long-standing hypotheses. We discuss the connection of the main sequence with one of the fundamental properties demonstrated by AGNs - Variability in Section \ref{sec3}, especially in advancing our knowledge through techniques like reverberation mapping (RM), and the renewed interest in Changing-look/Changing-state AGNs. We then give a brief account of the present-day scenario of incorporating AGNs (and quasars) as \textit{standard(izable)} candles and touch upon some relevant studies that have progressed in this direction. Finally, we conclude this mini-review with some closing remarks and perspective for the future in Section \ref{sec5} with up-and-coming massive, multiplex surveys that will make things more intriguing.

\section{Quasar Main Sequence - current state and advances}
\label{sec2}

In this section, we touch upon a few of the ongoing, interesting lines of research to improve our understanding of the main sequence of quasars. 

\subsection{Generating Fe II templates}

Owing to its complexity and uncertainties in transition probabilities and excitation mechanisms, the most successful approach to model the \feii{} emission in AGNs consists of deriving empirical templates from observations and supplementing the missing transitions with state-of-the-art radiative transfer models, e.g. CLOUDY \citep{Ferland_2017RMxAA, Chatzikos_etal_2023RMxAA}. The templates thus derived using this methodology are referred to as semi-empirical. {The work of \citet{Kovacevic_etal_2010ApJS} is seminal in this regard who provided the AGN community with an interface\footnote{\url{http://servo.aob.rs/FeII_AGN/}} to create \feii{} templates by combining the \feii{} transitions from theoretical expectations and those revealed in the spectrum of the prototypical \feii{}-emitter, I Zw 1. Their methodology involves the knowledge of the temperature of the ionized cloud responsible for the \feii{} emission, information on the dynamics of the \feii{} profile in the observed spectrum, and intensities of the strongest \feii{}  multiplets collected in three groups (b$^4$ F, a$^6$ S, and a$^4$ G). Their semi-empirical templates have been applied to large samples of AGNs, and demonstrated to provide convincing results}. All except one went back to the I Zw 1 - in the optical \citep[e.g.,][]{Veron-Cetty_etal_2004, marziani_etal_2021Univ}, in the UV \citep[e.g.,][]{Vestergaard_2001ApJS, Bruhweiler_2008ApJ, Tsuzuki_etal_2006ApJ}. The exception happened more recently with the HST/STIS observations for Mrk 493 \citep{Park_etal_2022ApJS} which has narrower lines, lower reddening, and a less extreme Eddington ratio value than I Zw 1, therefore, can be applied to a larger population of Type-1 AGNs with intrinsically lower \feii{} emission \citep{shen_ho_2014, panda_etal_2018ApJ}. {These templates also allow us to infer the velocity information of the \feii{} emission, a key aspect constraining the geometry and kinematics of the \feii{} emitting region in the BLR. Studies \citep{2008ApJ...687...78H, 2009ApJ...707L..82F, 2015ApJS..221...35K} have suggested that the \feii{} emission originates from a location different from, and most likely exterior to, the region that produces most of H$\beta$. These observational findings were confirmed through the analysis of emissivity profiles of AGNs using photoionization modeling \citep{panda_etal_2018ApJ, Sniegowska_etal_2020} and studies of \feii{} time-lags relative to the H$\beta$ in samples of AGNs using reverberation mapping \citep{Barth_etal_2013ApJ, Gaskell_etal_2022AN}.}

In recent years, there has been noteworthy development to improve the atomic datasets available for iron emission, with updated radiative and electron collisional rates, and include higher levels (up to 716) and energies as high as 26.4 eV. We refer the readers to \citet{Sarkar_etal_2021ApJ} for an overview of these datasets and their performance within the spectral synthesis code, CLOUDY.

\subsection{\feii{} spectral synthesis and inferring the BLR cloud properties}

Over the years after \citeauthor{boroson_green_1992} put forward their findings from the Eigenvector 1, the expected parameters that should influence the observed correlation in the quasar main sequence have been looked at, albeit separately. Notable among them are (i) the \feii{} emission model developed in \citet{Verner_1999ApJS} with 371 atomic levels producing 13,157 (permitted) emission lines with the highest energy level of $\sim$11.6 eV; (ii) study by \citet{Baldwin_etal_2004ApJ} which were among the first to suggest the importance of microturbulence ($\gtrsim$100 km s$^{-1}$)- the intra-cloud pressure broadening in the broad-line emitting region (BELR) clouds, to explain both the observed shape and equivalent width of the \feii{} emission. We note that the findings in \citeauthor{Baldwin_etal_2004ApJ} were confined to the UV regime to recover the 2200-2800 \AA\ \feii{} bump feature. They also suggested the need to include higher metal abundances in the BELR clouds to recover the observed \feii{} intensities, confirming the earlier results from observations by, e.g., \citet{Hamann_Ferland_1993ApJ, Hamann_Ferland_1999ARAA, Dietrich_etal_2003ApJ}; (iii) the need to have higher mean densities and column densities in the BELR clouds via numerical modeling to recover the \feii{} pseudocontinuum behavior, still in the UV regime, was suggested in the paper by \citeauthor{Bruhweiler_2008ApJ}, who re-affirmed the importance of the microturbulence in the BELR. 

In more recent years, a clearer picture of the \feii{} emission, especially in the optical region, linking to the quasar main sequence has been achieved. There is a growing consensus that the main sequence of quasars, earlier thought to be primarily driven by the Eddington ratio, is in reality, dependent on a combination of parameters of the underlying accretion disk and the BELR clouds \citet{panda_etal_2018ApJ, panda_etal_2019ApJ_WC, panda_etal_2019ApJ_QMS}. These parameters are: (1) Eddington ratio, (2) BH mass; (3) shape of the ionizing continuum (SED); (4) BLR density; (5) BLR metallicity; (6) velocity distribution of the BLR clouds (including microturbulence); (7) source's orientation; and (8) BLR cloud sizes \citep[see][]{Panda_2021PhDT}. The 8-dimensional parameter space was first presented by \citet{panda_etal_2019ApJ_QMS}, and extended by \citet{Panda_etal_2020CoSka} wherein through large grids of photoionization models with CLOUDY and massive observational spectroscopic catalogs \citep{shen_etal_2011, rakshit_etal_2020ApJS} the inherent trends along the main sequence have been confirmed. This almost completes the circle initiated with the hypotheses in \citet{boroson_green_1992} although more progress is needed, from observational and theoretical aspects. This multi-dimensional parameterization includes the viewing angle to the source (or orientation), which is constrained for a small fraction of the AGNs, especially those that show strong radio ``jetted'' emissions \citep[see e.g.,][ for an overview]{Padovani_etal_2017} or strong water masers \citep{Neufeld_1994ApJ, Greenhill_2003ApJ}. For the remaining sources, the viewing angle is estimated indirectly - through dynamical modeling \citep{Pancoast_2011ApJ, Yan-Rong_2013ApJ, Williams_etal_2018ApJ, Yan-Rong_2024arXiv}, through polarization studies of the emission lines \citep{Savic_etal_2018, Bo-Wei_2021MNRAS, sniegowska_pol_etal_2023, Jose_etal_2024MNRAS}, or broad-band SED modeling \citep{XCIGALE_2020MNRAS, AGNFitter_2024arXiv}. The knowledge of the viewing angle is crucial since it can be combined with the spatial and velocity distribution of BELR clouds and their location from the central ionizing source, to estimate the black hole mass of the source. The methodology presented by \citet{Panda_etal_2020CoSka} is powerful and acts in dual-purpose - for sources with known orientation and spectroscopically measured \feii{} emission, it can allow to constrain the BLR density and metallicity. On the other hand, through observed UV diagnostics if the BLR density and metallicity can be inferred (in addition to the \feii{} emission), one can recover the orientation angle of the source. The methodology, at present, includes the state-of-the-art broad-band SEDs presented in \citet{panda_etal_2019ApJ_QMS} and \citet{Ferland_etal_2020MNRAS}, while the BH mass, Eddington ratio, velocity distributions and line intensities are from the SDSS QSO catalogs \citep{shen_etal_2011, rakshit_etal_2020ApJS} for observed AGNs, and can be refined with future multi-wavelength campaigns. Recent works by \citet{pandey_etal_2023, pandey_etal_2024} have extended these results with new, and up-to-date \feii{} atomic datasets and accounting for dust within the BLR. Additionally, using these new datasets, \citet{ddds_etal_2023BoSAB, ddds_etal_2024Physi} have probed into the \feii{} emission in the NIR regime with the added advantage of transitions being isolated and less in number relative to the optical and UV.

In another recent work \citep{floris_etal_2024arxiv}, we performed a multi-component analysis on the strongest UV and optical emission lines and using $\sim$10 metal content diagnostic ratios that reveal a systematic progression in metallicity, ranging from sub-solar values to several times higher than solar values. This notable finding was a result of a series of papers \citep{sniegowska_etal_2021ApJ, garnica_etal_2022, marziani_etal_2024Physi} wherein a robust recipe of estimating metallicity and other physical parameters in highly-accreting Type-1 AGNs were developed. These results confirm the theoretical predictions made by \citet{panda_etal_2019ApJ_QMS} where the increase in the metal content was noted as a key factor that proportionally led to an increase in the \feii{} emission along the main sequence.

There are multiple studies predating the aforementioned papers that have paved the way to our current understanding of the \feii{} emission and we recommend the readers to the detailed accounts by e.g., \citet{sulentic_etal_2000, marziani_etal_2001ApJ, Zamfir_etal_2010MNRAS, 50Years_Book_2012ASSL, shen_ho_2014, Sulentic_etal_2015FrASS, marziani_etal_2018, Gaskell_etal_2022AN, panda_marziani_2023FrASS}.

\subsection{Developing AGN SEDs along the main sequence}

The shape of the ionizing continuum has been an integral part of the main sequence of quasars studies. Around the same time as \citeauthor{boroson_green_1992}, researchers were already developing mean AGN SEDs \citep{VandenBerk_2001AJ, Richards_etal_2006ApJS}, be it to distinguish the sources based on radio dichotomy \citep{Laor_etal_1997ApJ} and more recently in \citet{marziani_etal_2021Univ} or to reveal the prominence of the big blue bump feature in typical Type-1 AGNs \citep{Mathews_Ferland_1987ApJ, Korista_etal_1997ApJS}. With the advent of large spectroscopic surveys spearheaded by SDSS \citep{york_etal_2000AJ, shen_etal_2011} AGNs exhibiting stronger \feii{} emission alike I Zw 1 were being consistently discovered and led to the creation of a mean SED representing Narrow-line Seyfert 1 galaxies \citep{Marziani_Sulentic_2014MNRAS}. We now have broad-band mean SEDs grouped in Eddington ratios ranging from sub- to super-Eddington limits \citep{Jin_etal_2012MNRAS, Jin_etal_2017MNRAS, Ferland_etal_2020MNRAS}. Although these mean SEDs have helped provide statistical inferences on the role of AGN SED in the main sequence trends, having broad-band SED for individual AGNs is a much more recent endeavor that has seen growth. With the increase in simultaneous observations across multiple spectral regimes and the development of self-consistent AGN SED models \citep{done_etal_2012MNRAS, Kubota_Done_2018MNRAS, Kubota_Done_2019MNRAS, Hagen_2023MNRAS}, the number of individual sources with broad-band SEDs is growing at a rapid pace, especially for sources demonstrating the most intense \feii{} emission \citep{Marinello_etal_2020MNRAS, Jin_etal_2023MNRAS}. Another important extension in the area of SED building is the slim disk AGN SED models \citep{Abramowicz_1988ApJ...332..646A, Wang_etal_2014ApJ, panda_marziani_2023BoSAB} applicable to those sources accreting at or above the Eddington limit, that show signatures of strong outflows even in the low-ionization emitting regions \citep[e.g.,][]{rodriguez-ardila_etal_2024AJ}.  

\section{QMS and AGN variability}
\label{sec3}

Another equally important finding was the discovery of the variation in the intensities of emission lines over timescales of weeks to months, suggesting very small emitting regions of the order of a few thousand Schwarzschild radii \citep{Greenstein_Schmidt_1964ApJ}. This region is now well-known as the broad-line region (BLR). This crucial discovery opened up a new sub-field called reverberation mapping (RM), which has led to the estimation of black hole masses in hundreds of low- to high-luminosity Seyferts and quasars \citep{Blandford_Mckee_1982ApJ, Peterson_1988PASP, Peterson_1993PASP, Peterson_etal_2004ApJ}, supplemented by single/multi-epoch spectroscopy \citep{Kaspi_2000ApJ, Bentz_etal_2013ApJ, Du_etal_2016ApJ}. The BLR's location (R$_{\rm BLR}$) is closely related to the continuum properties of the underlying accretion disk, with luminosity being the primary observable quantity \citep[][and references therein]{Kaspi_etal_2005ApJ}. Subsequent studies, such as \citet{Bentz_etal_2013ApJ}, refined the H$\beta$-based R$_{\rm BLR}$ - L$_{\rm 5100}$ (or R-L) relation by including more sources and removing the host galaxy's contribution from the total luminosity. Increased monitoring of archival and newer sources has revealed a significant scatter from the empirical R-L relation \citep{Du_etal_2015ApJ, Grier_etal_2017ApJ, mlma_etal_2019ApJ, Du_Wang_2019ApJ, panda_etal_2019FrASS}. This scatter indicates a subset of sources with relatively high luminosities (log L$_{\rm 5100}$ = 43.0, in erg s$^{-1}$) that exhibit shorter time lags and thus shorter R$_{\rm BLR}$ than expected. Recent studies suggest that this scatter may be linked to the accretion rate, providing corrections to the empirical relation based on observables that trace the accretion rate, such as the strength of the optical \feii{} emission \citep{Du_Wang_2019ApJ, panda_2022FrASS, panda_marziani_2023FrASS}.

On the other hand, the complexity in the modeling and extracting \feii{} emission from the spectra has led many to search for viable alternatives. Most prominent among the proxy is the Ca {\sc ii} triplet (or CaT) in the NIR given the similarity of the physical conditions required to produce the two ionic species in the BLR \citep{panda_etal_2020ApJ_CaFe, panda_2021_CaFe}. In an ongoing series of works \citep{mlma_2015ApJS, Marinello_2016ApJ, panda_etal_2020ApJ_CaFe, mlma_etal_2021ApJ}, we have compiled optical \feii{} and NIR CaT emission strengths and weighed them against each other. We find a robust correlation between the two \citep{mlma_2015ApJS, panda_etal_2020ApJ_CaFe} primarily driven by the Eddington ratio and in parts to the BH mass \citep{mlma_etal_2021ApJ}. This led us to investigate whether CaT can be a viable replacement for the strength of the \feii{} emission (or \rfe{}) in the R-L relation \citep{mlma_etal_2021POBeo}. Although the current sample statistics are small in the NIR regime, the spurt of high-quality AGN spectra with the JWST and other ground-based facilities is promising. 

\subsection{Changing-Look AGNs and our renewed interest in them}

Changing-look AGNs have been known for almost as long as the main sequence existed \citep[see recent review by][]{Komossa_etal_2024arXiv, Ricci_Traktenbrot_2023NatAs}. The spectral changes over multiple epochs have now been detected in numerous AGNs - be it extreme variability with the changes in the continuum and emission lines so strong that can be associated with external interference such as obscuration or tidal disruption events \citep{LaMassa_etal_2015ApJ, Dodd_etal_2023ApJ, Benny_2019NatAs}, but could very well be associated with intrinsic effects such as disk transition/disk instabilities \citep{Noda_Done_2018MNRAS, Ross_etal_2018MNRAS, Sniegowska_etal_2020} although, the timescales of such events can be widely different \citep{Czerny_2006ASPC}.

With the growing interest in finding new changing-look AGNs, the focus has been also to look for AGNs showing variations in their \feii{} emission \citep[see e.g.,][]{Gaskell_etal_2022AN, Petrushevska_2023}. The regular variable nature of AGNs has helped to gain insights into their emitting regions, with some sources where we have estimates of their \feii{}-emitting locations \citep[see e.g.,][]{Hu_etal_2015ApJ, Barth_etal_2013ApJ} although there are now instances of exceptional changes in the \feii{} intensities. \citet{panda_sniegowska_2024} made a compilation of such sources and tracked their transition along the Eigenvector 1 schema and categorized sources that either stay within the same population (A or B, see right panel of Figure \ref{fig:enter-label}) or make an inter-population movement as a function of spectral epoch.

\subsection{New avenues in Reverberation mapping: BLR saturation and \feii{}-based R-L relations}

In addition to Changing-look AGNs, dedicated spectro-photometric monitoring campaigns on individual sources (e.g., Mrk 6, NGC 5548, NGC 4151, NGC 4051), have allowed us to re-affirm the Pronik-Chuvaev effect i.e., the increase, albeit with a gradual saturation, in the H$\beta$ emitting luminosity with increasing AGN continuum \citep{pronik_1972Ap, Wang_2005, Shapovalova_2008, Gaskell_2021MNRAS} and more recently in \citep{panda_etal_2022AN, panda_etal_2023BoSAB}. This assists in building the R-L relation for individual epochs and gaining insights into the temporal behavior of the line-emitting BLR relative to the continuum \citep{Zu_etal_2011ApJ, Li_etal_2022ApJS, 2024arXiv240901637F} {although studies to reveal the temporal behavior specifically in \feii{} are needed to complement the H$\beta$ behavior in these AGNs.}

Another interesting revelation has been the construction of the first \feii{} UV R-L relation in \citet{zajacek_etal_2024}. Here, in addition to improving the existing Mg {\sc ii}-based R-L with 194 sources (more recent compilation in \citealt{Shen_2024ApJS}), we have been able to constrain the R-L behavior in UV-emitting \feii{} for 5 AGNs. The results are motivating as the slope of the R-L is in close agreement with one expected from the standard photoionization theory (i.e., =0.5). Although it is interesting to note that this relation appears steeper than the Mg {\sc ii} R-L relations such that for low-luminosity regimes, the \feii{} emitting region is closer than the Mg {\sc ii}-emitting region, whereas, at higher luminosities, both relations converge and intersect. This intriguing behaviour needs more explanation which the upcoming RM campaigns may have an answer to. {We note that the optical \feii{}-based R-L relation has been around for some time \citep[see][for a recent review]{Gaskell_etal_2022AN}. A recent compilation of 17 AGNs (including multiple epoch \feii{} time lag measurements) from \citet{prince_etal_2023} reveals an R-L for the optical \feii{} with a slope close to 0.5, and the comparison with the aforementioned UV-based R-L reveals an offset by a factor of 1.8, i.e., the optical \feii{} emitting regions are located 1.8 times further out relative to the UV \feii{} regions.}

\subsection{Quasars for cosmology: role of the main sequence}
\label{sec4}

Quasars, with their extragalactic origin and persistent bright nature, have long been proposed as ``standardizable candles’’. With the knowledge of their luminosities (with the aid of the RM and R-L relation) and independently of their fluxes from spectroscopic monitoring, we can determine the luminosity distances of these sources. With a growing number of AGNs (now $\gtrsim$200, \citealt{zajacek_etal_2024, Shen_2024ApJS}) where we have such estimates, then allows us to prepare a Hubble diagram - stretching the redshift regime to higher ranges as compared to what we can achieve with other indicators, e.g., Cepheids, Tip of the Red Giant Branch (TRGB), and Type-1a Supernovae (SNIa). We can then scrutinize the various, existing cosmological models, in addition to gauging the performance of quasars to existing indicators, allowing us to link the cosmological measurements from the early Universe (e.g. \citealt{Planck_2020}) to the measurements from the late Universe (e.g. Cepheids, SNIa, and TRGBs; \citealt{Riess_1998AJ, Riess_2019ApJ, Freedmann_2019ApJ}).

However, the recent detection of shorter lags in R-L linked to high-accreting sources \citep{Du_etal_2015ApJ, Grier_etal_2017ApJ, Du_etal_2016ApJ} has put the use of R-L relation into uncertainty. In \citet{mlma_etal_2019ApJ}, we looked into the dispersion in the R-L and upon further investigation found the extent of offset of the source's time-lag is proportional to the Eddington ratio (or more specifically its mass accretion rate). While this helped ``standardize'' the R-L relation, there remained a circularity problem - the mass accretion rate needs the knowledge of luminosity apriori, and the latter can be estimated assuming a cosmological model. This defeats the purpose of using quasars for cosmology and thus, requires us to find a direct observable parameter that can replace the mass accretion rate. What can be that? In \citet{Du_Wang_2019ApJ}, the authors found that the \feii{} strength (or \rfe{}) is a viable alternative to the mass accretion rate (as has been noted in earlier works of \citealt{marziani_etal_2018, panda_etal_2019ApJ_QMS}) and can correct the dispersion in the R-L. This \rfe{}-dependent R-L relation has hence been tested and confirmed in other works \citep{panda_2022FrASS, panda_marziani_2023FrASS}. Other empirical relations notably the L$_{\rm X}$ - L$_{\rm UV}$ relation, are proposed as a viable alternative to the R-L relation \citep{Risaliti_Lusso_2015ApJ, Risaliti_Lusso_2019NatAs} although there are subtle differences between the two relations and their inferences. Yet another methodology has been proposed, i.e., with the aid of the existing correlation between the luminosity and the velocity distribution of the BELR \citep{Dultzin_2020FrASS, Marziani_2021IAUS}, equivalent to the original formulation of the Faber-Jackson law \citep{Faber-Jackson_1976ApJ}. We refer the readers to \citet{panda_marziani_2023FrASS} for more details.

In a parallel direction, efforts to reconcile the use of quasars along with other distance indicators, e.g, SNIa, Gamma-ray bursts, Baryon Acoustic Oscillations, and temperature anisotropy across the microwave background, have been made including the \rfe{} parameter for the quasar-based R-L relations \citep{Cao_2022MNRAS, Khadka_2023MNRAS, Dainotti_2023ApJ}. Other systematics, such as dust extinction can contribute to and reconcile the observed difference between the R-L and L$_{\rm X}$ - L$_{\rm UV}$ relation \citep{zajacek_2024ApJ}. {As we enter into the discussions around Hubble-Lema\^itre law and the H$_{\rm 0}$ tension, a key to resolving this is better measurements of cosmological distances. VLT/GRAVITY has opened up avenues to probe the angular sizes of the BLR in nearby AGNs using high-resolution spectroastrometry \citep{GRAVITY_2018Natur, GRAVITY_2020, GRAVITY_2021, GRAVITY_2024_RL}. These angular sizes can be combined with the BLR linear sizes (the latter estimated using the RM technique) to give the parallax distance to these AGNs. The technique was originally conceived in \citet{Elvis_2002ApJ} although thanks to the recent interferometric measurements by GRAVITY coupled with their long-term RM monitoring campaign, \citet{Wang_SARM_2020NatAs} have been able to estimate, for the first time, a H$_{\rm 0}$ value using this joint analysis with AGNs. Ongoing improvements with GRAVITY \citep[see e.g.,][]{GRAVITY_plus_2024} will allow the compilation of a sizable sample of AGNs extending to z $\sim$ 2 where the spectroastrometric-RM (or, SARM) technique can be applied to build the Hubble diagram for quasars. This however requires the knowledge of the BLR properties that are neatly tied to the quasar main sequence which positively affect the accuracy of the estimation of the cosmological distances to these cosmic objects.}

\section{Closing remarks and future perspective}
\label{sec5}

\feii{} emission has been long perceived as a contaminant in the AGN spectra and ways to remove this contamination were sought to enable study and reliable extraction of other emission line properties. The emission turned out to be so useful that a niche of studies linking to the \feii{} emission was proposed and expanded. To date, the studies stemming from the \feii{} analysis have crucial contributions in developing our understanding of the line-emitting regions in the BLR leading up to the standardization of quasar-based scaling relations. This mini-review cannot do justice to the enormous literature about the study of \feii{} emission and its link to the quasar main sequence. Yet, we have tried to touch upon some key aspects in this short overview. We are already in the data-driven astronomy era with multiple facilities working in cohesion, to reveal more connections to the quasar main sequence.

Finally, we glance upon some recent avenues that will have a direct impact on the ongoing studies:

\begin{itemize}
    \item Ongoing and upcoming spectroscopic surveys such as JWST \citep{JWST_2023PASP}, MSE \citep{MSE_2019BAAS}, WST \citep{WST_2024arXiv}, 4MOST \citep{4MOST_2019Msngr} are going to help reveal intriguing \feii{} signatures in low-luminosity regimes and distant quasars, e.g., JWST ASPIRE \citep[and references therein][]{ASPIRE_2023ApJ} showing high \feii{} emitting AGNs beyond the cosmic noon; putting into question the prevalence of heavy metals in such early epochs. {Additionally, Additionally, the first couple of years of observations with JWST has revealed the numerous faint, broad-line AGN at z $>$ 5 \citep{2023ApJ...942L..17O, 2023ApJ...954L...4K, 2023ApJ...959...39H, 2024ApJ...963..129M, 2023arXiv230801230M, 2023ApJ...953L..29L, 2024ApJ...964...39G}. A significant fraction of them ($\sim$20\%) show a steep red continuum in the rest-frame optical region, in addition to being relatively bluer in the UV \citep{2023ApJ...954L...4K, 2023ApJ...959...39H, 2024ApJ...963..129M, 2024ApJ...964...39G, 2023arXiv231203065K} giving the appearance of a ``V-shape'' in the SEDs for these intriguing objects. These sources, also known as ``little red dots'' \citep[LRDs,][]{2024ApJ...963..129M, 2024arXiv240403576K}, and while the prominent broad emission lines (i.e., Balmer lines) are relatively easier to deblend. Their profiles can be fitted even under moderate spectral quality, while complex emissions like the \feii{} would require future deeper observations to check their location on the main sequence of quasars to reveal their nature and chemical history at such redshifts.}

    \item On the other hand, the large-scale photometric surveys (e.g., ZTF: \citealt{ZTF_2019PASP}, LSST: \citealt{LSST_2019ApJ}, Euclid: \citealt{Euclid_2022}) will identify newer AGNs, and combined with the wide-area spectroscopic surveys will allow constraining the \feii{} contribution and improve our understanding of the various mechanisms involved in \feii{} production across UV-optical-NIR regime, especially dealing with time-lag recovery and extracting broad-band SED for tens of hundreds of AGNs across a wide range of redshifts \citep[see][for a recent review]{panda_2023Univ}.

    \item With the LSST about to begin its decade-long survey, the use of meter-class ground-based facilities in cohesion with such massive surveys will be pertinent \citep{Chelouche_2019NatAs, Panda_2024ApJL}; narrow-band filters will allow optimizing the lag-recovery by mitigating spectral windows with contamination. While, the use of traditional and machine-learning techniques are going to be integral for target selection from erstwhile surveys \citep{Baron_2019arXiv, Sanchez-Saez_2021AJ, lopez-navas_2022MNRAS, sniegowska_pol_etal_2023}.
\end{itemize}

\section*{Conflict of Interest Statement}

The author declares that the research was conducted in the absence of any commercial or financial relationships that could be construed as a potential conflict of interest.

\section*{Author Contributions}

SP conceived the idea, designed the layout, collected the data, and wrote and edited the manuscript.


\section*{Acknowledgments}
SP acknowledges the financial support of the Conselho Nacional de Desenvolvimento Científico e Tecnológico (CNPq) Fellowships 300936/2023-0 and 301628/2024-6. SP is supported by the international Gemini Observatory, a program of NSF NOIRLab, which is managed by the Association of Universities for Research in Astronomy (AURA) under a cooperative agreement with the U.S. National Science Foundation, on behalf of the Gemini partnership of Argentina, Brazil, Canada, Chile, the Republic of Korea, and the United States of America. This mini-review has been made possible thanks to many past and ongoing collaborations; I would like to thank Bo\.zena Czerny, Paola Marziani, Alberto Rodr\'iguez Ardila, Mary Loli Mart\'inez-Aldama, Murilo Marinello, Marzena \'Sniegowska, Francisco Pozo-Nu\~nez, Michal Zaja\v{c}ek, Edi and Nata\v{s}a Bon, Szymon Koz\l{}owski and many others for their invaluable support, constant motivation and fruitful discussions. I am grateful to the organizers of the ``Frontiers in Astronomy and Space Sciences: A Decade of Discovery and Advancement - 10th Anniversary Conference'' for the invitation to write this mini-review.



\bibliographystyle{Frontiers-Harvard} 
\bibliography{frontiers_revised}



\end{document}